\documentclass[aps,prl,twocolumn,groupedaddress,showpacs,floatfix]{revtex4}
\usepackage{dcolumn}
\usepackage{amsmath}
\usepackage{amssymb}
\usepackage{graphicx}
\usepackage{bm}
\usepackage[T1]{fontenc}
\usepackage{color}
\usepackage{url}
\usepackage[bf]{subfigure}
\usepackage{rotating}

\begin{document}

\title{A Complex Networks Approach for Data Clustering}

\author{Francisco A. Rodrigues}
\email{francisco@icmc.usp.br}
\affiliation{Departamento de Matemática Aplicada e Estatística, Instituto de Ciências Matemáticas e de Computação,
Universidade de São Paulo - Campus de São Carlos, Caixa Postal 668,
13560-970 São Carlos, SP.}
\author{Guilherme Ferraz de Arruda}
\author{Luciano da Fontoura Costa}
\affiliation{Instituto de F\'{\i}sica
de S\~{a}o Carlos, Universidade de S\~{a}o Paulo, Av. Trabalhador
S\~{a}o Carlense 400, Caixa Postal 369, CEP 13560-970, S\~{a}o
Carlos, S\~ao Paulo, Brazil}

\begin{abstract}
Many methods have been developed for data clustering, such as k-means,
expectation maximization and algorithms based on graph theory. In this
latter case, graphs are generally constructed by taking into account
the Euclidian distance as a similarity measure, and partitioned using
spectral methods. However, these methods are not accurate when the
clusters are not well separated. In addition, it is not possible to
automatically determine the number of clusters. These limitations can
be overcome by taking into account network community identification
algorithms. In this work, we propose a methodology for data clustering
based on complex networks theory. We compare different metrics for
quantifying the similarity between objects and take into account three
community finding techniques. This approach is applied to two
real-world databases and to two sets of artificially generated
data. By comparing our method with traditional clustering approaches,
we verify that the proximity measures given by the Chebyshev and
Manhattan distances are the most suitable metrics to quantify the
similarity between objects. In addition, the community identification
method based on the greedy optimization provides the smallest
misclassification rates.
\end{abstract}


\maketitle

\section{Introduction}

Classification is one of the most intrinsic activities of human
beings, being used to facilitate the handling and organization of the
huge amount of information that we receive every day. As a matter of
fact, the brain is able to recognize objects in scenes and also to
provide a categorization of objects, persons, or events.  This
classification is performed in order to cluster objects that are
similar with respect to common attributes. Actually, humans have by
now classified almost all known living species and materials on
earth. Due to the importance of the classification task, it is
fundamental to develop methods able to perform this task
automatically. Indeed, many methods for categorization have been
developed with application to life sciences (biology, zoology),
medical sciences (psychiatry, pathology), social sciences (sociology,
archaeology), earth sciences (geography, geology), and
engineering~\cite{Anderberg,Jain99}.

The process of classification can be performed in two different ways,
\emph{i.e.} \emph{supervised classification}, where the previously known
class of objects are provided as prototypes for classifying additional
objects; and \emph{unsupervised classification}, where no previous
knowledge about the classes is provided. In the latter case, the
categorization is performed in order to maximize the similarity
between the objects in each class while minimizing the similarity
between objects in different classes. In the current work, we
introduce a method for unsupervised classification based on complex
networks.

Unsupervised classification may be found under different names in
different contexts, such as clustering (in pattern recognition),
numerical taxonomy (in ecology) and partition (in graph theory). In
the current work, we adopt the term ``clustering''. Clustering can be
used in many tasks, such as \emph{data reduction}, performed by
grouping data into cluster and processing each cluster as a single
entity; \emph{hypothesis generation}, when there is no information
about the analyzed data; \emph{hypothesis testing}, \emph{i.e.}
verification of the validity of a particular hypothesis; and
\emph{prediction based on classes}, where the obtained clusters are
based on the characteristics of the respective patterns. As a matter
of fact, clustering is a fundamental tool for many research fields,
such as machine learning, data mining, pattern recognition, image
analysis, information retrieval, and bioinformatics~\cite{Jain99,
Everitt}.

Many methods have been developed for data
clustering~\cite{theodoridispattern}, many of which are based on graph
theory~\cite{Jain1999}. Graphs-based clustering methods take into
account algorithms related to minimum spanning trees~\cite{Zahn06},
region of influence (e.g.~\cite{Urquhart1982}), direct
trees~\cite{Koontz06} and spectral analysis~\cite{Koontz06}. These
methods are able to detect clusters of various shapes, at least for
the case in which they are well separated. However, these algorithms
present some drawbacks, such as the spectral clustering, which only
divides the graph into two groups and not in an arbitrary number of
clusters. Division into more than two groups can be achieved by
repeated bisection, but there is no guarantee of reaching the best
division into three groups~\cite{theodoridispattern}. Also, these
methods give no hint about how many clusters should be identified. On
the other hand, methods for community identification in networks are
able to handle these drawbacks~\cite{Newman04PRE}. Moreover, these
methods provide more accurate partitions than the traditional method
based on graph, such as the spectral
partition~\cite{Newman04PRE}. Actually, methods based on complex
networks are improvements of clustering approaches based on graphs.

Only recently, a method has been developed for data clustering based
on complex networks concepts~\cite{Oliveira08}.  In this case, the
authors proposed a clustering method based on graph partitioning and
the Chameleon algorithm~\cite{Karypis02}. Although this method is able
to detect clusters in different shapes, it presents some
drawbacks. The authors considered a method for community
identification very particular which does not provide the most
accurate network division~\cite{Fortunato10}. In addition, it
considered only a single metric to establish the connections between
every pair of objects, \emph{i.e.} the Euclidian distance. On the
other hand, the method introduced in the current work overcomes all
these limitations. We adopt the most accurate community identification
methods and use the most traditional metrics to define the similarity
between objects, including the Euclidian, Manhattan, Chebyshev, Fu and
Tanimoto distances~\cite{theodoridispattern}. The accuracy of our methodology is
evaluated in artificial as well as two real-world databases. Moreover,
we compare our methodology with some traditional clustering
algorithms, i.e.\ k-means, cobweb, expectation maximization and
farthest first. We verify that our approach provides the smallest
error rates. So, we concluded that complex networks theory seems to
provide the tools and concepts able to improve the clustering methods
based on graphs, potentially overcoming the most traditional
clustering methods.

\section{Concepts and Methods}

\subsection{Complex networks}

Complex networks are graphs with non-trivial topological features,
whose connections are distributed as a
power-law~\cite{Barabasi:survey}. An undirected network can be
represented by its adjacency matrix $A$, whose elements $a_{ij}$ are
equal to one whenever there is a connection between the vertices $i$
and $j$, or equal to zero otherwise. A more general representation
takes into account weighted connections, where each edge $(i,j)$
presents an associated weight or strength $\omega(i,j)$.

Different measures have been developed to characterize the topology of
network structures, such as the clustering coefficient,
distance-related measurements and centrality
metrics~\cite{Costa:survey}. By allowing the different network
properties to be quantified, these methods have revealed that most
real-world networks are far from purely random~\cite{Newman10}.

In addition to this highly intricate topological organization, complex
networks also tend to present modular structure. In this case, these
modules are clusters whose vertices present similar roles, such as in
the case of the brain of mammals, where cortical modules are
associated to brain functions~\cite{Bullmore2009}. Communities have
the same principle as clusters in pattern recognition research. In
this way, the algorithms developed for community identification can
also be used to partition graph and finding clusters.

Different methods have been developed in order to find communities in
networks. Basically, these methods can be grouped as spectral methods
(e.g.~\cite{Newman0:PNAS}), divisive methods
(e.g.~\cite{Girvan02:PNAS}), agglomerative methods
(e.g.~\cite{Clauset:04PRE}), and local methods
(e.g.~\cite{Clauset:2005}). The choice of the best method depends of
the specific application, including the network size and number of
connections. This is due to the fact that the most precise methods,
such as the extremal optimization algorithm, are quite time
expensive. Here, we take three different methods that provide accurate
results, but have different time complexities. These methods are
described in the next section.

The quality of a particular network division can be evaluated in terms
of the \emph{modularity measure}. This metric allows the number of
communities to be automatically determined according to the best
network partition. For a network partitioned into $m$ communities, a
matrix $E$, $c \times c$, is constructed whose elements $e_{ij}$,
represent the fraction of connections between communities $i$ and
$j$. The modularity $Q$ is calculated as
\begin{equation}\label{Eq:modularity}
     Q = \sum_i [ e_{ii} - ( \sum_j e_{ij} )^2 ] = \mathrm{Tr} E  - ||E^2||.
\end{equation}
The highest value of modularity is obtained for the best network
division. In particular, networks that present high values of $Q$ have
modular structure implying that clusters are identified with high
accuracy~\cite{Newman04PRE, Newman10}.

\subsection{Clustering based on network}

In literature, there are many definitions of
clusters~\cite{theodoridispattern}, such as that provided by Everitt
\emph{et al.}~\cite{Everitt}, where clusters are understood as continuous
regions of the feature space containing a high density of points,
separated from other high density regions by low density regions. This
definition is similar to that of network communities,
\emph{i.e.} a community is topologically defined as a subset of highly
inter-connected vertices which are relatively sparsely connected to
nodes in other communities~\cite{Fortunato10}.

Let each object (also denominated pattern) be represented by a feature
vector $\vec{x} = [x_1,x_2,\ldots,x_n]$. These features, $x_i$, are
scalar numbers and quantify the properties of objects. For instance,
in case of the Iris database, the objects are flowers and the
attributes are the length and the width of the sepal and the petal, in
centimeters~\cite{Fisher1936}. The clustering approach consists of
grouping the feature vectors into $m$ clusters, $C_1,C_2,\ldots,C_m$,
in such a way that objects belonging to the same cluster exhibit
higher similarity with each other than with objects in other groups.

The process of clustering based on networks involves the definition of
the following concepts:
\begin{enumerate}
\item \emph{Proximity measure}: each object is represented as a node,
where each pair of nodes are connected according to their
similarity. These connections are weighted in the sense so as to
quantify how similar each pair of vertices is, in terms of
their feature vector. In this way, the most similar objects are
connected by the strongest edges.
\item \emph{Clustering criterion}: modularity is the most traditional
measure used to quantify de quality of a network
division~\cite{Fortunato10}, see Equation~\ref{Eq:modularity}. Here,
we adopt this metric to automatically choose the best cluster
partition. In problems in which the number of clusters is known, it is
not necessary to consider the modularity.
\item \emph{Clustering
algorithms}: Complex networks theory provides many algorithms for
community identification, which act as the clustering
algorithms~\cite{Fortunato10}. The choice of the most suitable method
for a particular application should take into account the error rate
and the execution time.
\item \emph{Validation of the results}:
The validation of the clustering methods based on networks can be
performed in two different ways: (i) by considering databases in which
the clusters are known (or at least expected), such as the Iris
database~\cite{Fisher1936}, and (ii) by taking into account artificial
data with cluster organization, which allows to control the level of
the data modular organization.
\end{enumerate}

Proximity measures can be classified into two types, \emph{similarity
measures}, that is $s(\vec{x},\vec{y}) = s_0$ only if $\vec{x}=\vec{y}$
and $-\infty<s(\vec{x},\vec{y})\leq s_0 < +\infty$; and \emph{dissimilarity
measures}, where $d(\vec{x},\vec{y}) = d_0$ only if $\vec{x}=\vec{y}$
and $-\infty<d_0 \leq d(\vec{x},\vec{y})< +\infty$. To construct
networks, it is more natural to adopt similarity measures, since it
is expected that the edges with the strongest weights should be
verified between the vertices with the most similar feature
vectors. In this way, we adopt the following similarity measures to
develop the network-based clustering
approach~\cite{theodoridispattern}:
\begin{enumerate}
 \item Inverse of Euclidian distance:
\begin{equation}
 D_E^{-1}(\textbf{x}, \textbf{y}) = \frac{1}{d_2(\textbf{x}, \textbf{y})},
\end{equation}
where $d_2(\textbf{x}, \textbf{y})$ is the traditional Euclidian distance.
This metric results in values in the interval $[0, \infty)$.
 \item Exponential of Euclidian distance:
\begin{equation}
 S_E(\textbf{x}, \textbf{y}) = \alpha \exp \left(-\alpha d_2(\textbf{x}, \textbf{y}) \right),
\end{equation}
where this metric results in values in the interval $[0, \alpha]$.
\item Inverse of Manhattan distance:
\begin{equation}
 D_M^{-1}(\textbf{x}, \textbf{y}) = \left(\sum_{i=1}^n |x_i-y_i|\right)^{-1},
\end{equation}
which assumes values in $[0, \infty)$.
\item Exponential of Manhattan distance:
\begin{equation}
 S_{M}(\textbf{x}, \textbf{y}) = \alpha \exp \left( -\alpha D_M \right),
\end{equation}
assuming values in the interval $[0, \alpha]$.
\item Inverse of Chebyshev distance:
\begin{equation}
 D_C^{-1}(\textbf{x}, \textbf{y}) = \frac{1}{\max_{i=1}^n |x_i-y_i|}.
\end{equation}
This metric results in values in $[0, \infty)$.
\item Exponential of Chebyshev distance:
\begin{equation}
 S_{C}(\textbf{x}, \textbf{y}) = \alpha \exp \left( -\alpha D_C \right),
\end{equation}
assuming values in the interval $[0, \alpha]$.
 \item Metric proposed by Fu,
\begin{equation}
 F(\textbf{x}, \textbf{y}) = 1 - \frac{d_2(\textbf{x}, \textbf{y})}{\parallel\textbf{x}\parallel+\parallel\textbf{y}\parallel}.
\end{equation}
This metrics results in values in the interval $[0, 1]$
 \item Exponential of the metric proposed by Fu:
\begin{equation}
S_F(\textbf{x}, \textbf{y}) = \alpha \exp \left( -\alpha \frac{1 -  F}{2} \right).
\end{equation}
If $F(\textbf{x}, \textbf{y})=1$, then $S_F(\textbf{x}, \textbf{y}) = \alpha$. If  $F(\textbf{x}, \textbf{y})=0$, then
$S_F(\textbf{x}, \textbf{y}) = \alpha \exp \left(\frac{- \alpha}{2} \right)$. Therefore, $S_F$ assumes values in this limited interval.
 \item Exponential of the Tanimoto mesure:
\begin{equation}
S_T = \alpha \exp \left( -\alpha \frac{1 -  T}{2} \right),
\end{equation}
where
\begin{equation}
 T(\textbf{x}, \textbf{y}) = \frac{\textbf{x}^T\textbf{y}}{{\parallel\textbf{x}\parallel}^2 + {\parallel\textbf{y}\parallel}^2 - \textbf{x}^T\textbf{y}}.
\end{equation}
This metric assumes values in $(-\infty, 1]$. Therefore, if $T(\textbf{x}, \textbf{y})=1$, then, $s_T(\textbf{x}, \textbf{y}) = \alpha$, if $T(\textbf{x}, \textbf{y}) \rightarrow - \infty$, then,
$S_T(\textbf{x}, \textbf{y}) \rightarrow 0$
\end{enumerate}

In order to divide networks into communities and therefore obtain the
clusters, we adopt three methods, namely the maximization of the
modularity method, which is based on the greedy
algorithm~\cite{Clauset:04PRE}, here called \emph{fastgreedy}
algorithm; the extremal optimization approach~\cite{Duch:2005} and the
\emph{waltrap} method~\cite{Pons05:CIS}. In both former methods,
two communities $i$ and $j$ are joined according to the increase of
the modularity $Q$ of the network. Thus, starting with each vertex
disconnected and considering each of them as a community, we
repeatedly join communities together into pairs, choosing at each step
the merging that results in the greatest increase (or smallest
decrease) of the modularity $Q$. The best division corresponds to the
partition that resulted in the highest value of $Q$. The difference
between these two methods lies in the choice of the optimization
algorithm. On the other hand, the walktrap method is based on random
walks, where the community identification uses a metrics that
considers the probability transition matrix~\cite{Pons05:CIS}. The
time of execution of the walktrap method run as $O(N^2\log N)$. While
the fastgreedy method is believed to be the fastest one, running in
$O(N\log^2 N)$, the extremal optimization provides the most accurate
division~\cite{Danon:2005}. On the other hand, the extremal
optimization method is not particularly fast, scaling as $O(N^2\log
N)$.

\begin{figure*}[!tpb]
\centerline{\includegraphics[width=0.9\linewidth]{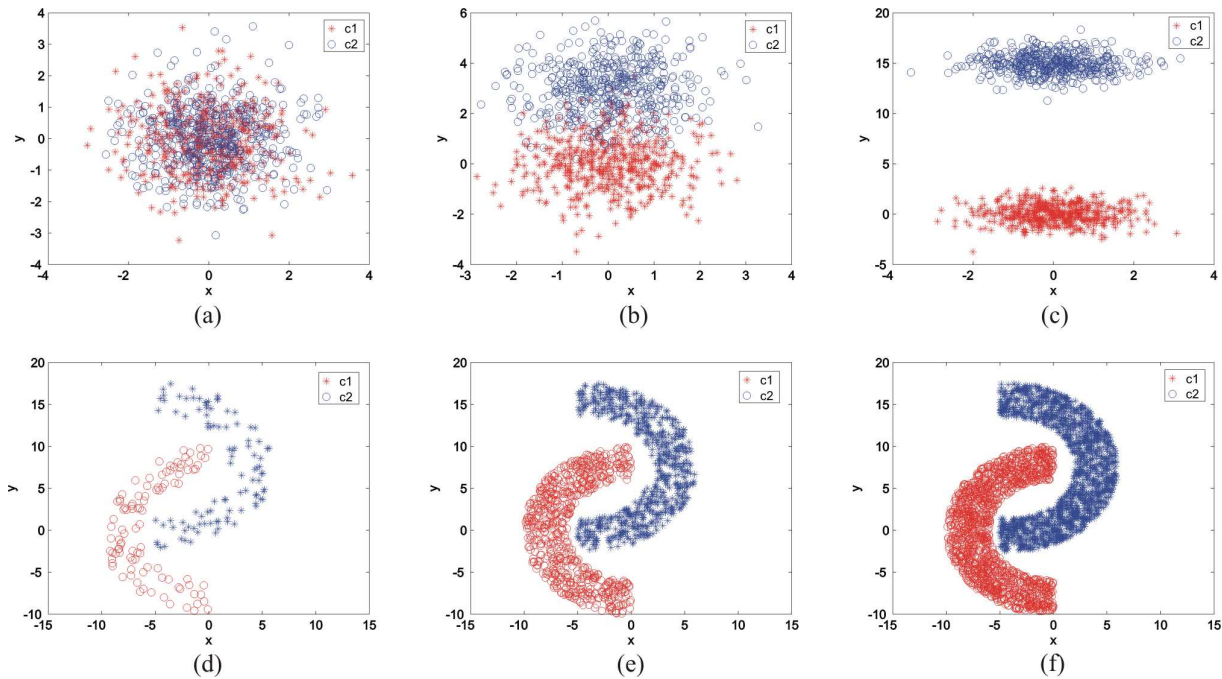}}
   \caption{Examples of the artificial databases for clustering
   evaluation error. The first set points is generated according to a
   Gaussian distribution, where the two sets are separated by
   distances (a) $d=0$, (b) $d = 3$ and (c) $d = 15$. In the case of
   the second set, a higher density of points provides more defined
   clusters, as shown in examples (d) $\rho = 1.0$, (e) $\rho = 6.4$
   and (f) $\rho = 14.4$.} \label{Fig:bases}
\end{figure*}

The validation of the network-based clustering method is performed
with respect to artificial (\emph{i.e.} computer generated clusters)
and real-world databases. In the case of artificial data, we use two
different configurations, \emph{i.e.} (i) two separated clouds of
points with a Gaussian distribution in a two dimensional space, and
(ii) two semi-circles with varying density of points, as presented in
Figure~\ref{Fig:bases}. In the former case, the validation set
consists of two set of points (clusters) generated according to a
gaussian distribution with covariance matrix equals to identity,
($\Sigma = I$). The median of one set of points is moved from the
origin $(0,0)$ until (0,15), in steps of 0.75, while the other cluster
remains fixed at the origin of axis. In this way, the distance between
clusters is varied from $d=0$ to $d=15$. Figure~\ref{Fig:bases}(a) to
(c) shown three cases considering three distances, \emph{i.e.} $d=0$,
$3$ and $15$. Observe that as $d$ increases, the cluster
identification becomes easier. The second artificial database
corresponds to a classic problem in pattern
recognition~\cite{theodoridispattern}. It consists of two sets of
points uniformly generated in two limited semi-circle areas. In this
case, the density of points, \emph{i.e.} the number of points by unit
of area, defines the cluster resolutions, with higher density
producing more defined clusters. In our analysis, this density is
varied from 1 to 32, in steps of 1.6. Figures~\ref{Fig:bases}(d) to
(f) show three configurations of this artificial database generated by
taking into account three different densities, $\rho = 1$, 6.4 and
14.4.

With respect to real-world databases, we take into account two
datasets, \emph{i.e.} the Iris database~\cite{Fisher1936}, and the
Breast Cancer Wisconsin database~\cite{Wolberg94}. The Iris database
is composed by three species of Iris flowers (\emph{Iris setosa},
\emph{Iris virginica} and \emph{Iris versicolor}). Each class consists
of 50 samples, where four features were measured from each sample,
\emph{i.e.} the length and the width of the sepal and the petal, in
centimeters. On the other hand, the cancer database is composed by
features of digitized image of a fine needle aspirate from a breast
mass, where 30 real-valued features are computed for each cell
nucleus~\cite{Wolberg94}. This database is composed by 699 cells,
where 241 are malignant and 458 are benign.

\section{Results and discussion}

The accuracy of the clustering method based on networks is compared
with four traditional clustering methods, namely k-means, cobweb,
farthest first and expectation maximization
(EM)~\cite{Witten05}. These methods present different properties, such
as the k-means tendency to find spherical
clusters~\cite{theodoridispattern}. Moreover, we consider three
methods for community identification, namely fastgreedy, extremal
optimization and walktrap~\cite{Fortunato10}. However, in this work,
since the fastgreedy and extremal optimization result in the same
error rates for all considered databases, we discuss only the results
of the fastgreedy method, which is faster than the extremal
optimization approach.

\begin{figure}[!t]
\begin{center}
\subfigure[]{\includegraphics[width=0.9\linewidth]{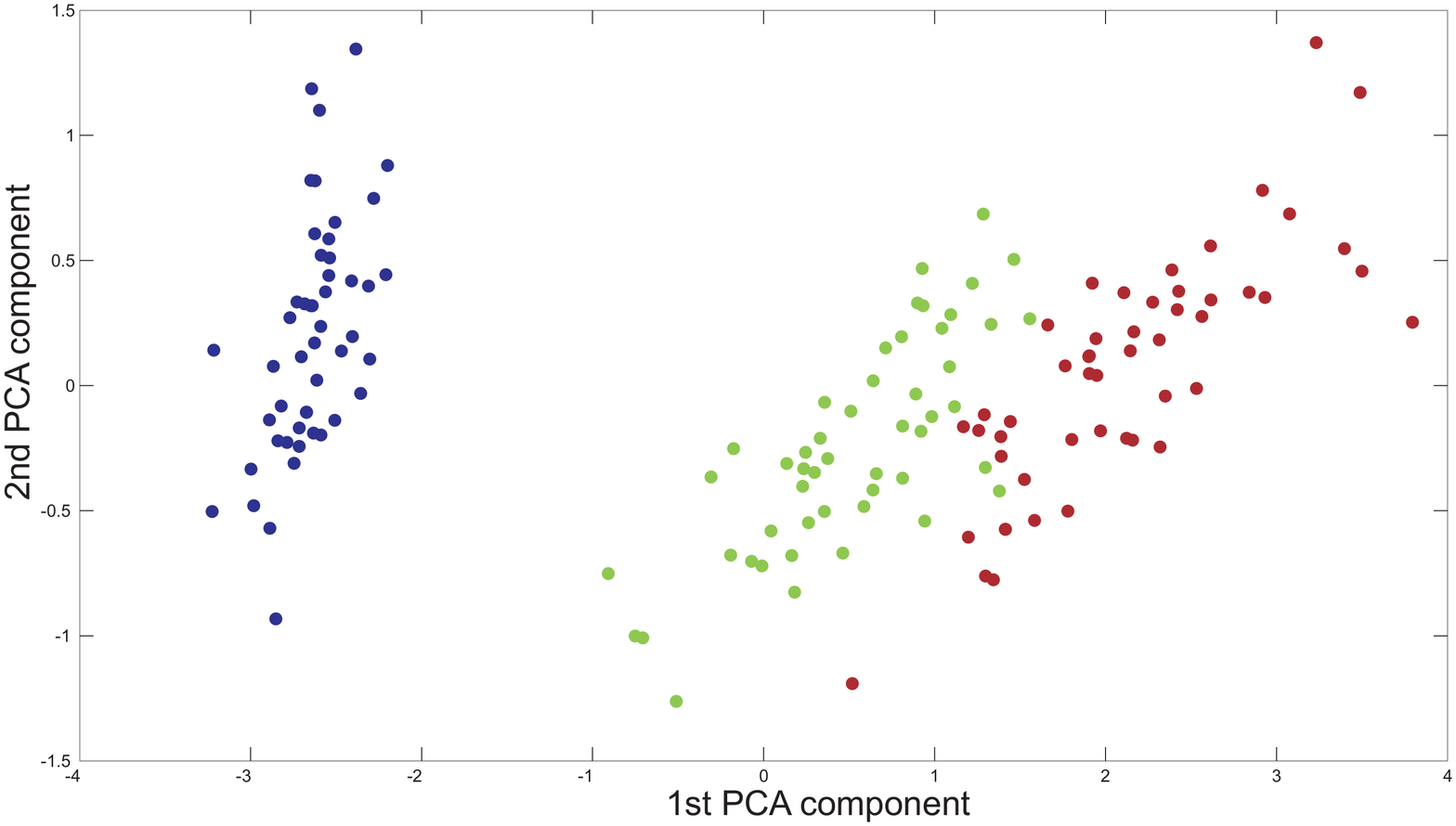}} \\
\subfigure[]{\includegraphics[width=0.9\linewidth]{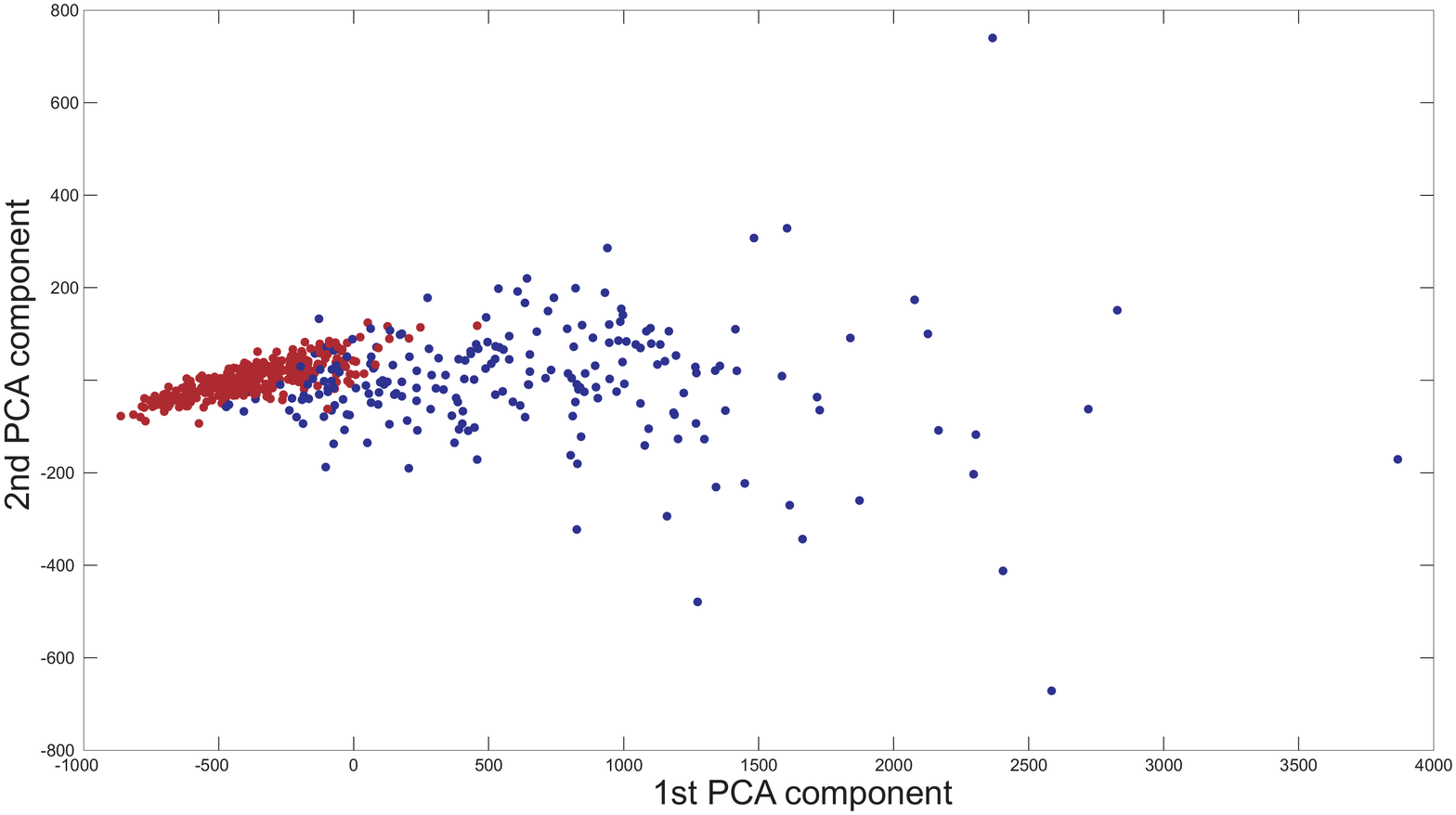}}
\end{center}
\caption{Projection of (a) the Iris and (b) the Breast Cancer Wisconsin
databases by principal component analysis.}
\label{Fig:pca}
\end{figure}

We start our analysis by taking into account the Iris and the Breast
Cancer Wisconsin databases. As a preliminary data visualization, we
project the patterns into a two dimensional space by taking into
account principal component analysis. Figure~\ref{Fig:pca} shows the
projections. It is clear that there is no clear separation between the
clusters for both databases.

Since the attributes in the Iris data present different ranges, having
values such as 0.1 for the petal width and 7.2 for the sepal length,
it is necessary to take into account a feature standardization
procedure~\cite{theodoridispattern}. In this case, each attribute is
transformed in order to present mean equals to zero and standard
deviation equals to one. This transformation, called standardization,
is performed as,
\begin{equation}
y_f = \frac{x_f - \overline{x_f}}{\sigma_{x_f}}
\end{equation}
where $\overline{x_f}$, $\sigma_{x_f}$ are the average and standard
deviation of the values of attribute $f$, respectively. The obtained
results considering the four clustering algorithm is presented in
Table~\ref{Tab:iris}. The EM and k-means exhibit the smaller errors
among the traditional classifiers. However, note that this performance
is obtained when the number of clusters is known. On the other hand,
EM provides an error of 40\% when the number of clusters is
unknown. This is a limitation of these methods, since in most of the
cases, the information about the number of classes is not available.

\begin{figure}[!t]
\centerline{\includegraphics[width=1\linewidth]{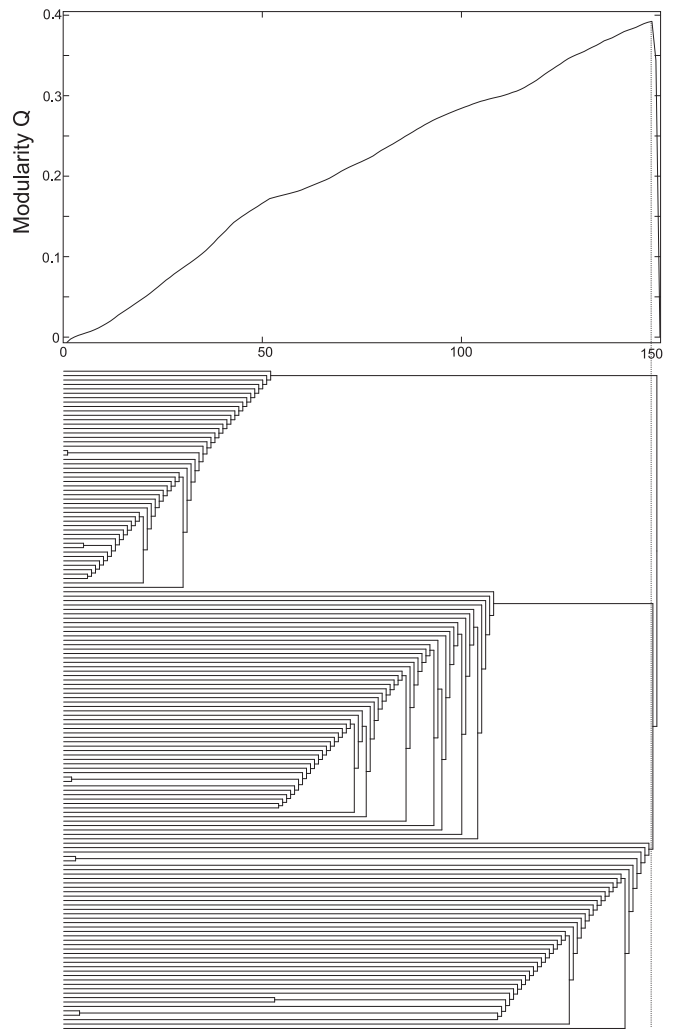}}
\caption{Dendrogram obtained by taking into account the inverse of
Chebyshev distance and the fastgreedy community identification
algorithm for the Iris data.  The cut in the dendrogram results in
three classes, where the error rate is equal to 4.7\%.}
\label{Fig:dendrogram}
\end{figure}

Table~\ref{Tab:iris} also presents the results with respect to the
cluster-based on complex networks approaches. Only combinations
between metric and community algorithm which result in the smallest
error rates are shown in this table. The smallest error was obtained
by taking into account the inverse of the Chebyshev distance and the
fastgreedy community identification algorithm. The obtained error for
this case is equal to 4.7\%. The second best performance is obtained
by considering the inverse of the Euclidian distance and the
fastgreedy or the walktrap algorithms, which provide an error of
6\%. In addition to the smallest error rates, network-based clustering
present other important feature, \emph{i.e.} it is not necessary to
specify the number of clusters present in the database. Indeed, the
maximum value of the modularity suggests the most accurate
partition. Nevertheless, for some proximity measures, the modularity
is not able to determine the best partition. In this case, the
knowledge about the number of clusters implies in a reduction of the
error rates, as in the case of exponential of the Tanimoto distance,
where the error is reduced from 33.3\% to 6\%, and the exponential of
the Chebyshev distance, where the error is reduced from 33.3\% to
7.3\%. Therefore, for the Iris data, such metrics are not appropriated
for network-based clustering. We also analyze the clustering error
without standardization. In this case, the error rates are larger than
those obtained considering the normalization, for some cases. However,
for the best results, we verify that the errors are similar in both
cases. Figure~\ref{Fig:dendrogram} presents the dendrogram obtained
for the best separation, \emph{i.e.} by taking into account the
inverse of Chebyshev distance and the fastgreedy community
identification method. Observe that the best partition is obtained for
the highest value of the modularity measure.

\begin{center}
\begin{table}[!t]
\centering
\caption{Clustering errors for the Iris database considering the cases in which the number of classes $k$ is known ($k=3$) or unknown ($k = ?$).
EM and k-means are the only methods that need to specify the number of
clusters $k$.} \label{Tab:iris}
\begin{tabular}{|l|c|c|}
\hline
Method & \% error ($k=?$) & \% error ($k = 3$) \\
\hline
k-means & -- &11.3\\
\hline
cobweb & 33.3 & --\\
\hline
farthest first & -- & 14.0 \\
\hline
EM & 40.0 & 9.3\\
\hline
$D_E^{-1}$ - fastgreedy & 6.0 & 6.0\\
\hline
$D_E^{-1}$ - walktrap & 6.0 & 6.0\\
\hline
$S_E$ - walktrap & 33.3 & 14.7\\
\hline
$S_T$ - walktrap & 33.3  & 6.0\\
\hline
$S_F$ - walktrap & 33.3 & 7.3 \\
\hline
$D_M^{-1}$ - fastgreedy & 33.3 & 6.0\\
\hline
$D_M^{-1}$ - walktrap & 33.3 & 6.0\\
\hline
$S_M$ - walktrap & 33.3 & 6.0\\
\hline
$D_C^{-1}$ - fastgreedy & 4.7 & 4.7 \\
\hline
$D_C^{-1}$ - walktrap & 9.3 & 9.3 \\
\hline
$S_C$ - Walktrap & 33.3 & 7.3 \\
\hline
\end{tabular}
\end{table}
\end{center}

The cancer database also needs to be pre-processed by the
standardization. The obtained clustering errors are presented in
Table~\ref{Tab:cancer}. Only combinations between metric and community
algorithm which result in the smallest error rates are shown in this
table. In this case, the smallest clustering error is obtained by the
k-means method, which produces an error rate of 7.2\%. However, the
complex networks-based method taking into account the inverse of
Manhattan distance and walktrap algorithm for community identification
provides an error rate of 7.9\%. Observe that when the number of
clusters is known, all methods result in smaller error
rates. Nevertheless, the $D_M^{-1}$-walktrap produces the same error
rate of 7.9\% even when $k=2$. Therefore, the highest value of the
modularity accounts for the separation for this method. Although our
proposed method implied in an higher error than the k-means
methodology, it presents the advantage that it is not necessary to
known the number of clusters. In this way, our methodology is also
more suitable to determine the clusters for the Breast Cancer
Wisconsin database.

\begin{center}
\begin{table}[!t]
\centering
\caption{Clustering errors for the Breast Cancer Wisconsin database
considering the cases in which the number of classes $k$ is known
($k=3$) or unknown ($k = ?$). EM and k-means are the only methods that
need to specify the number of clusters $k$.} \label{Tab:cancer}
\begin{tabular}{|l|c|c|}
\hline
Method & \% error (k=?) & \% error (k = 3) \\
\hline
k-means &-- &7.2\\
\hline
cobweb & 37.2 &--\\
\hline
farthest first & -- & 35.3 \\
\hline
EM  & 75.9 & 8.8\\
\hline
$D_E^{-1}$ - walktrap & 52.9 & 9.8\\
\hline
$S_F$ - fastgreedy & 17.6 & 17.6 \\
\hline
$D_M^{-1}$ - walktrap & 7.9 & 7.9\\
\hline
$D_C^{-1}$ - fastgreedy & 50.8 & 15.3 \\
\hline
$S_C$ - walktrap & 15.9 & 15.9 \\
\hline
\end{tabular}
\end{table}
\end{center}

\begin{figure}[!t]
\centerline{\includegraphics[width=1\linewidth]{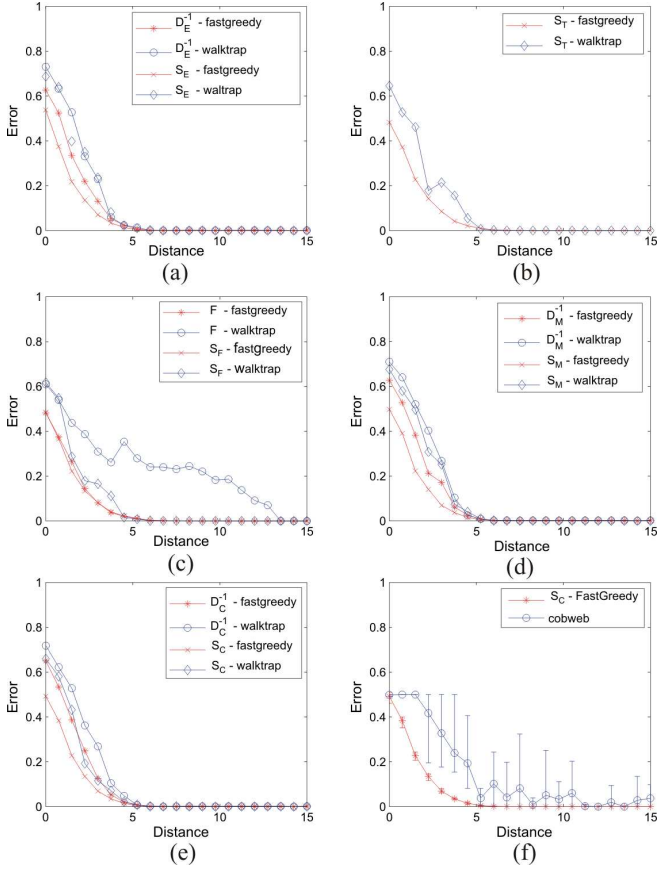}}
\caption{Error rates obtained according to the separation of two clusters
composed by sets of points with gaussian distribution in a
two-dimensional space. We show the best results for each proximity
measure, i.e.\ (a) Euclidian, (b) Tanimoto, (c) Fu, (d) Manhattan and
(e) Chebyshev distances. The number of clusters is determined
automatically by the maximum value of the modularity for all cases. In
(f) the most accurate network-based method is compared
with the best traditional clustering approach. Each point is an
average over 10 simulations.}
\label{Fig:gaussian_kauto}
\end{figure}

In order to provide a more comprehensive evaluation of the proposed
complex networks-based clustering approach, we generated two set of
artificial data into a two dimensional space, as discussed in the last
section. This artificial data allows to control the cluster
separability of the generated databases. Initially, we consider two
clusters of points with Gaussian distribution in an two-dimensional
space separated by a distance $d$. Figure~\ref{Fig:gaussian_kauto}
presents the best obtained results for the complex networks-based
approach taking into account different proximity measures. For all
cases, the number of clusters is determined automatically by the
maximum value of the modularity. Note that the error rate goes to zero
for $d \geq 5$. Figure~\ref{Fig:gaussian_kauto}(f) shows the
comparison between the traditional clustering method which resulted in
the best results, \emph{i.e.} the cobweb, and the best complex
networks approach. In this case, the method based on the exponential
of the Chebyshev distance and fastgreedy algorithm provides the
smallest error rate. Observe that the variation of the error rate is
also small for this method, compared with the cobweb.

\begin{figure}[!t]
\centerline{\includegraphics[width=1\linewidth]{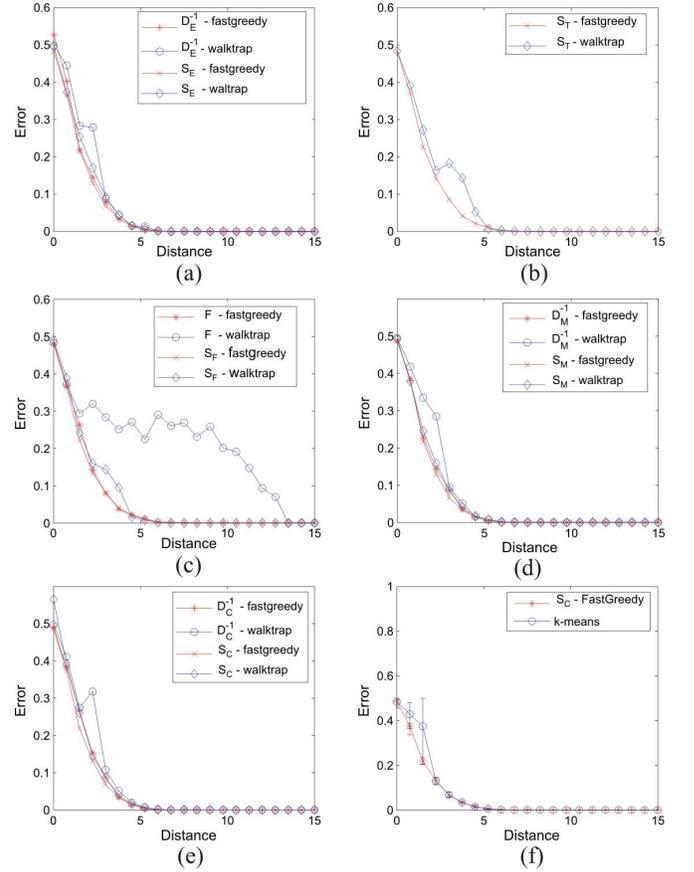}}
\caption{Error rates obtained according to the separation of two clusters
composed by sets of points with Gaussian distribution in a
two-dimensional space for the case where the number of clusters is
known. We show the best results for each proximity measure,
\emph{i.e.} (a) Euclidian, (b) Tanimoto, (c) Fu, (d) Manhattan and (e)
Chebyshev distances. The number of clusters is set as $k=2$ for all
methods. In (f) it the most accurate network-based method is compared
with the best traditional clustering approach. Each
point is an average over 10 simulations.}
\label{Fig:gaussian_kfix}
\end{figure}

The k-means algorithm cannot be used in the comparison where the
number of clusters $k$ is known. Thus, we consider the case where $k$
is determined for all methods. Figure~\ref{Fig:gaussian_kfix} presents
the obtained results. In all cases, the error rate goes to zero for $d
\geq 5$. As in the case of unknown number of clusters, the method
based on the exponential of the Chebyshev distance and fastgreedy
algorithm provides the smallest error rate. Among the traditional
algorithms, the k-means allows the most accurate results. In fact, for
$d \geq 3$, both k-means and network-based clustering method provide
similar error rates. For this database, accurate results were expected
for the k-means method, since the clusters are symmetric around the
means, being equally distributed among the two clusters. Observe that
the other approaches of the complex networks-based method also imply
in an small error rate. Comparing with the cases where the number of
clusters is unknown, \emph{i.e.}  Figure~\ref{Fig:gaussian_kauto}
shows that the error rate is similar for both approaches. Therefore,
the network-based methods result in the most accurate cluster
partitions and have the advantage that it is not necessary to know the
number of clusters.

\begin{figure}[!t]
\centerline{\includegraphics[width=0.9\linewidth]{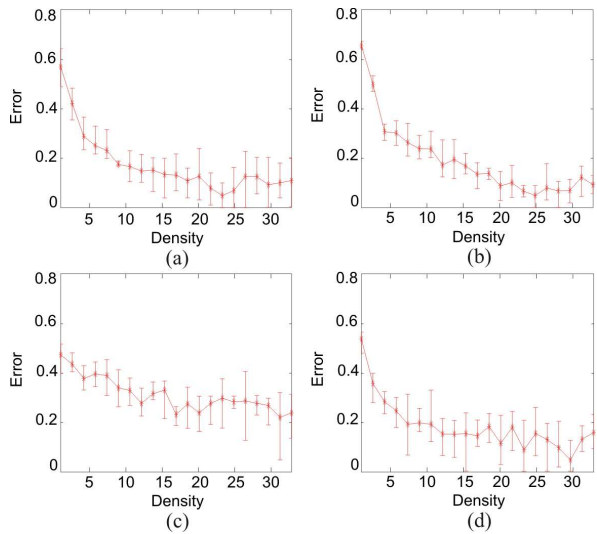}}
\caption{The smallest clustering errors obtained for the complex
networks-based methods applied in the second artificial dataset (see
Figures~\ref{Fig:bases}(d) -- (f)). The error rates are determined
according to the density of points. The adopted proximity measures are
(a) the exponential of the Euclidian distance, (b) the exponential of
the Manhattan distance, (c) the inverse of the Chebyshev distance and
(d) the exponential of the Chebyshev distance. The number of clusters
is obtained automatically by the maximum value of the modularity.  }
\label{Fig:art_kauto}
\end{figure}

\begin{figure}[!t]
\centerline{\includegraphics[width=0.9\linewidth]{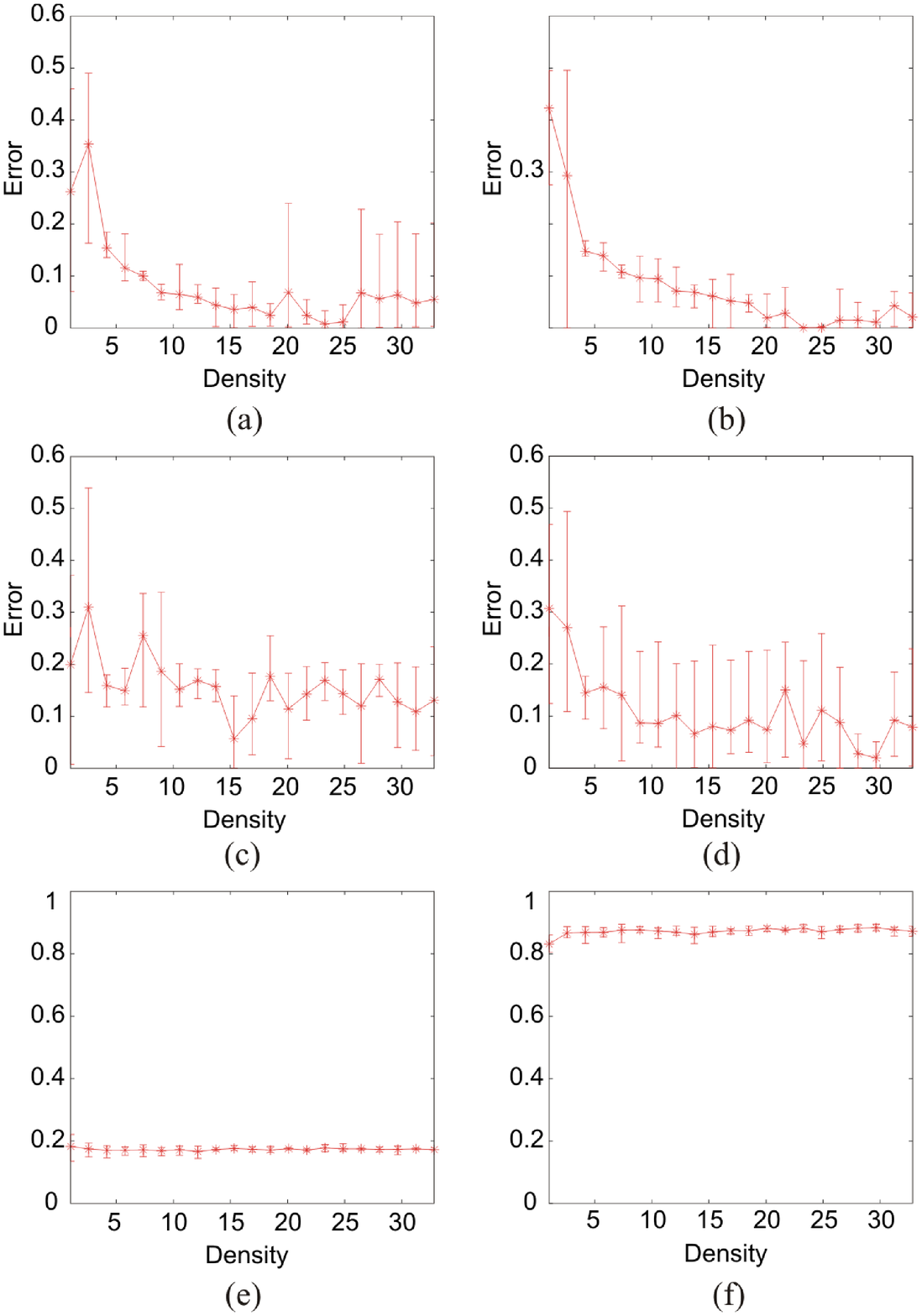}}
\caption{The smallest clustering errors obtained for the complex
networks-based methods applied in the second artificial dataset (see
Figures~\ref{Fig:bases}(d) -- (f)). The error rates are determined
according to the density of points. The adopted proximity measures are
(a) the exponential of the Euclidian distance, (b) the exponential of
the Manhattan distance, (c) the inverse of the Chebyshev distance and
(d) the exponential of the Chebyshev distance. The number of clusters
is fixed as $k=2$. The k-means (e) and cobweb (f) are the traditional
clustering methods that produce the smallest errors.}
\label{Fig:Res_art_kfixo}
\end{figure}

\begin{figure}[!t]
\centerline{\includegraphics[width=1\linewidth]{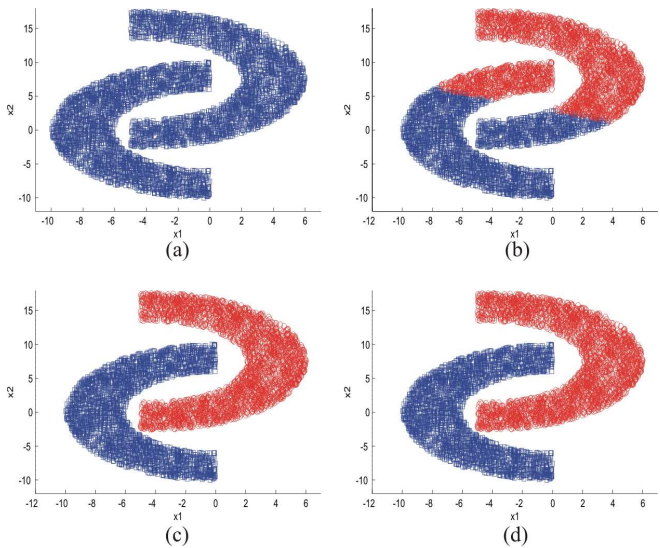}}
\caption{Example of the best performance for the (b) k-means, complex
network-based methods method using (c) the best modularity value and
(d) fixing $k=2$. The original data is shown in (a).}
\label{Fig:ex}
\end{figure}

The second artificial database used to evaluate the classification
error rates is given by the two semi-circle with varying density of
points (see Figures~\ref{Fig:bases}(d) --
(f)). Figure~\ref{Fig:art_kauto} presents the obtained results for
unknown number of clusters considering fastgreedy algorithm. Only the
best results are shown in this figure. The higher the density of
points, the smaller the error rate, since the clusters become more
defined. The error rate does not tend to zeros only for the inverse of
the Chebyshed distance (Figure~\ref{Fig:art_kauto}(c)). The most
accurate clustering is obtained by taking into account the exponential
of the Manhattan distance
(Figure~\ref{Fig:art_kauto}(b)). Figure~\ref{Fig:Res_art_kfixo}
presents the obtained errors when the number of clusters is known,
\emph{i.e.} $k=2$. Again, the complex networks-based method which
takes into account the exponential of the Manhattan distance produces
the smallest error. It is interesting to note that the traditional
clustering methods, i.e.\ k-means and cobweb, result in higher error
rates than the methods based on complex networks. In addition, the
error does not tend to zero when the density of points is increased
for these traditional methods. Figure~\ref{Fig:ex} presents an example
of the best clustering for the k-means and complex networks-based
methods. Observe that k-means cannot identify the correct clusters.

\section{Conclusion}

In this work, we study different proximity measures to represent a
data set into a graph and then adopt community detection algorithms to
perform respective clustering. Our obtained results suggest that
complex networks theory has tools to improve graph-based clustering
methodologies, since this new area provides more accurate algorithms
for community identification. In fact, comparing with traditional
clustering methods, the network-based approach finds clusters with the
smallest error rates for both real-world and artificial databases. In
addition, this methodology allows the identification of the number of
clusters automatically by taking into account the maximum value of the
modularity measurement. Among the considered proximity measures, the
inverse of the Chebyshev distance and the inverse of the Manhattan
distance are the most suitable metric for the considered real-world
databases. With respect to the artificial databases, the exponential
of the Chebyshev and exponential of the Manhattan distance produces
the smallest error rates. Therefore, metrics based on the Chebyshed
and Manhattan distances are the most suitable to quantify the
similarity between objects in terms of their feature vectors. Among
the community identification algorithms, the fastgreedy revealed to be
the most suitable, due to its accuracy and the smallest time for
processing.

The analysis proposed in this work can be extended by taking into
account other real-world databases as well as other approaches to
generate artificial clusters. The application to different areas, such
as medicine, biology, physics and economy constitute other promising
research possibilities.

\section*{Acknowledgement}
Luciano da F. Costa thanks CNPq (301303/06-1) and FAPESP (05/00587-5)
for sponsorship.

\bibliographystyle{apsrev}
\bibliography{paper}

\begin{thebibliography}{26}
\expandafter\ifx\csname natexlab\endcsname\relax\def\natexlab#1{#1}\fi
\expandafter\ifx\csname bibnamefont\endcsname\relax
  \def\bibnamefont#1{#1}\fi
\expandafter\ifx\csname bibfnamefont\endcsname\relax
  \def\bibfnamefont#1{#1}\fi
\expandafter\ifx\csname citenamefont\endcsname\relax
  \def\citenamefont#1{#1}\fi
\expandafter\ifx\csname url\endcsname\relax
  \def\url#1{\texttt{#1}}\fi
\expandafter\ifx\csname urlprefix\endcsname\relax\def\urlprefix{URL }\fi
\providecommand{\bibinfo}[2]{#2}
\providecommand{\eprint}[2][]{\url{#2}}

\bibitem[{\citenamefont{Anderberg}(1973)}]{Anderberg}
\bibinfo{author}{\bibfnamefont{M.}~\bibnamefont{Anderberg}},
  \emph{\bibinfo{title}{{Cluster Analysis for Applications}}}
  (\bibinfo{year}{1973}).

\bibitem[{\citenamefont{Jain et~al.}(1999{\natexlab{a}})\citenamefont{Jain,
  Murty, and Flynn}}]{Jain99}
\bibinfo{author}{\bibfnamefont{A.}~\bibnamefont{Jain}},
  \bibinfo{author}{\bibfnamefont{M.}~\bibnamefont{Murty}}, \bibnamefont{and}
  \bibinfo{author}{\bibfnamefont{P.}~\bibnamefont{Flynn}},
  \bibinfo{journal}{ACM Computing Surveys} \textbf{\bibinfo{volume}{31}},
  \bibinfo{pages}{264} (\bibinfo{year}{1999}{\natexlab{a}}), ISSN
  \bibinfo{issn}{0360-0300}.

\bibitem[{\citenamefont{Everitt et~al.}(2001)\citenamefont{Everitt, Landau, and
  Leese}}]{Everitt}
\bibinfo{author}{\bibfnamefont{B.~S.} \bibnamefont{Everitt}},
  \bibinfo{author}{\bibfnamefont{S.}~\bibnamefont{Landau}}, \bibnamefont{and}
  \bibinfo{author}{\bibfnamefont{M.}~\bibnamefont{Leese}},
  \emph{\bibinfo{title}{{Cluster analysis}}} (\bibinfo{publisher}{Arnold},
  \bibinfo{year}{2001}).

\bibitem[{\citenamefont{Theodoridis and
  Koutroumbas}(2003)}]{theodoridispattern}
\bibinfo{author}{\bibfnamefont{S.}~\bibnamefont{Theodoridis}} \bibnamefont{and}
  \bibinfo{author}{\bibfnamefont{K.}~\bibnamefont{Koutroumbas}},
  \emph{\bibinfo{title}{{Pattern recognition}}} (\bibinfo{publisher}{Academic
  Press}, \bibinfo{year}{2003}).

\bibitem[{\citenamefont{Jain et~al.}(1999{\natexlab{b}})\citenamefont{Jain,
  Murty, and Flynn}}]{Jain1999}
\bibinfo{author}{\bibfnamefont{A.}~\bibnamefont{Jain}},
  \bibinfo{author}{\bibfnamefont{M.}~\bibnamefont{Murty}}, \bibnamefont{and}
  \bibinfo{author}{\bibfnamefont{P.}~\bibnamefont{Flynn}},
  \bibinfo{journal}{ACM computing surveys (CSUR)}
  \textbf{\bibinfo{volume}{31}}, \bibinfo{pages}{264}
  (\bibinfo{year}{1999}{\natexlab{b}}), ISSN \bibinfo{issn}{0360-0300}.

\bibitem[{\citenamefont{Zahn}(2006)}]{Zahn06}
\bibinfo{author}{\bibfnamefont{C.}~\bibnamefont{Zahn}},
  \bibinfo{journal}{Computers, IEEE Transactions on}
  \textbf{\bibinfo{volume}{100}}, \bibinfo{pages}{68} (\bibinfo{year}{2006}),
  ISSN \bibinfo{issn}{0018-9340}.

\bibitem[{\citenamefont{Urquhart}(1982)}]{Urquhart1982}
\bibinfo{author}{\bibfnamefont{R.}~\bibnamefont{Urquhart}},
  \bibinfo{journal}{Pattern recognition} \textbf{\bibinfo{volume}{15}},
  \bibinfo{pages}{173} (\bibinfo{year}{1982}), ISSN \bibinfo{issn}{0031-3203}.

\bibitem[{\citenamefont{Koontz et~al.}(2006)\citenamefont{Koontz, Narendra, and
  Fukunaga}}]{Koontz06}
\bibinfo{author}{\bibfnamefont{W.}~\bibnamefont{Koontz}},
  \bibinfo{author}{\bibfnamefont{P.}~\bibnamefont{Narendra}}, \bibnamefont{and}
  \bibinfo{author}{\bibfnamefont{K.}~\bibnamefont{Fukunaga}},
  \bibinfo{journal}{Computers, IEEE Transactions on}
  \textbf{\bibinfo{volume}{100}}, \bibinfo{pages}{936} (\bibinfo{year}{2006}),
  ISSN \bibinfo{issn}{0018-9340}.

\bibitem[{\citenamefont{Newman and Girvan}(2004)}]{Newman04PRE}
\bibinfo{author}{\bibfnamefont{M.}~\bibnamefont{Newman}} \bibnamefont{and}
  \bibinfo{author}{\bibfnamefont{M.}~\bibnamefont{Girvan}},
  \bibinfo{journal}{Physical review E} \textbf{\bibinfo{volume}{69}},
  \bibinfo{pages}{26113} (\bibinfo{year}{2004}), ISSN
  \bibinfo{issn}{1550-2376}.

\bibitem[{\citenamefont{de~Oliveira et~al.}(2008)\citenamefont{de~Oliveira,
  Zhao, Faceli, and de~Carvalho}}]{Oliveira08}
\bibinfo{author}{\bibfnamefont{T.}~\bibnamefont{de~Oliveira}},
  \bibinfo{author}{\bibfnamefont{L.}~\bibnamefont{Zhao}},
  \bibinfo{author}{\bibfnamefont{K.}~\bibnamefont{Faceli}}, \bibnamefont{and}
  \bibinfo{author}{\bibfnamefont{A.}~\bibnamefont{de~Carvalho}}, in
  \emph{\bibinfo{booktitle}{IEEE Congress on Evolutionary Computation, 2008.
  CEC 2008.(IEEE World Congress on Computational Intelligence)}}
  (\bibinfo{year}{2008}), pp. \bibinfo{pages}{2121--2126}.

\bibitem[{\citenamefont{Karypis et~al.}(2002)\citenamefont{Karypis, Han, and
  Kumar}}]{Karypis02}
\bibinfo{author}{\bibfnamefont{G.}~\bibnamefont{Karypis}},
  \bibinfo{author}{\bibfnamefont{E.}~\bibnamefont{Han}}, \bibnamefont{and}
  \bibinfo{author}{\bibfnamefont{V.}~\bibnamefont{Kumar}},
  \bibinfo{journal}{IEEE Computer} \textbf{\bibinfo{volume}{32}},
  \bibinfo{pages}{68} (\bibinfo{year}{2002}), ISSN \bibinfo{issn}{0018-9162}.

\bibitem[{\citenamefont{Fortunato}(2010)}]{Fortunato10}
\bibinfo{author}{\bibfnamefont{S.}~\bibnamefont{Fortunato}},
  \bibinfo{journal}{Physics Reports} \textbf{\bibinfo{volume}{486}},
  \bibinfo{pages}{75} (\bibinfo{year}{2010}), ISSN \bibinfo{issn}{0370-1573}.

\bibitem[{\citenamefont{Albert and Barab\'{a}si}(2002)}]{Barabasi:survey}
\bibinfo{author}{\bibfnamefont{R.}~\bibnamefont{Albert}} \bibnamefont{and}
  \bibinfo{author}{\bibfnamefont{A.-L.} \bibnamefont{Barab\'{a}si}},
  \bibinfo{journal}{Reviews of Modern Physics} \textbf{\bibinfo{volume}{74}},
  \bibinfo{pages}{48} (\bibinfo{year}{2002}).

\bibitem[{\citenamefont{da~F.~Costa et~al.}(2007)\citenamefont{da~F.~Costa,
  Rodrigues, Travieso, and Boas}}]{Costa:survey}
\bibinfo{author}{\bibfnamefont{L.}~\bibnamefont{da~F.~Costa}},
  \bibinfo{author}{\bibfnamefont{F.~A.} \bibnamefont{Rodrigues}},
  \bibinfo{author}{\bibfnamefont{G.}~\bibnamefont{Travieso}}, \bibnamefont{and}
  \bibinfo{author}{\bibfnamefont{P.~R.~V.} \bibnamefont{Boas}},
  \bibinfo{journal}{Advances in Physics} \textbf{\bibinfo{volume}{56}},
  \bibinfo{pages}{167 } (\bibinfo{year}{2007}).

\bibitem[{\citenamefont{Newman}(2010)}]{Newman10}
\bibinfo{author}{\bibfnamefont{M.~E.~J.} \bibnamefont{Newman}},
  \emph{\bibinfo{title}{{Networks: An Introduction}}}
  (\bibinfo{publisher}{Oxford Univ Pr}, \bibinfo{year}{2010}), ISBN
  \bibinfo{isbn}{0199206651}.

\bibitem[{\citenamefont{Bullmore and Sporns}(2009)}]{Bullmore2009}
\bibinfo{author}{\bibfnamefont{E.}~\bibnamefont{Bullmore}} \bibnamefont{and}
  \bibinfo{author}{\bibfnamefont{O.}~\bibnamefont{Sporns}},
  \bibinfo{journal}{Nature Reviews Neuroscience} \textbf{\bibinfo{volume}{10}},
  \bibinfo{pages}{186} (\bibinfo{year}{2009}), ISSN \bibinfo{issn}{1471-003X}.

\bibitem[{\citenamefont{Newman}(2006)}]{Newman0:PNAS}
\bibinfo{author}{\bibfnamefont{M.}~\bibnamefont{Newman}},
  \bibinfo{journal}{Proceedings of the National Academy of Sciences}
  \textbf{\bibinfo{volume}{103}}, \bibinfo{pages}{8577} (\bibinfo{year}{2006}).

\bibitem[{\citenamefont{Girvan and Newman}(2002)}]{Girvan02:PNAS}
\bibinfo{author}{\bibfnamefont{M.}~\bibnamefont{Girvan}} \bibnamefont{and}
  \bibinfo{author}{\bibfnamefont{M.}~\bibnamefont{Newman}},
  \bibinfo{journal}{Proceedings of the National Academy of Sciences of the
  United States of America} \textbf{\bibinfo{volume}{99}},
  \bibinfo{pages}{7821} (\bibinfo{year}{2002}).

\bibitem[{\citenamefont{Clauset et~al.}(2004)\citenamefont{Clauset, Newman, and
  Moore}}]{Clauset:04PRE}
\bibinfo{author}{\bibfnamefont{A.}~\bibnamefont{Clauset}},
  \bibinfo{author}{\bibfnamefont{M.~E.~J.} \bibnamefont{Newman}},
  \bibnamefont{and} \bibinfo{author}{\bibfnamefont{C.}~\bibnamefont{Moore}},
  \bibinfo{journal}{Physical Review E} \textbf{\bibinfo{volume}{70}},
  \bibinfo{pages}{066111} (\bibinfo{year}{2004}).

\bibitem[{\citenamefont{Clauset}(2005)}]{Clauset:2005}
\bibinfo{author}{\bibfnamefont{A.}~\bibnamefont{Clauset}},
  \bibinfo{journal}{Physical Review E} \textbf{\bibinfo{volume}{72}},
  \bibinfo{pages}{026132} (\bibinfo{year}{2005}).

\bibitem[{\citenamefont{Fisher}(1936)}]{Fisher1936}
\bibinfo{author}{\bibfnamefont{R.~A.} \bibnamefont{Fisher}},
  \bibinfo{journal}{Annals Eugenics} \textbf{\bibinfo{volume}{7}},
  \bibinfo{pages}{12} (\bibinfo{year}{1936}).

\bibitem[{\citenamefont{Duch and Arenas}(2005)}]{Duch:2005}
\bibinfo{author}{\bibfnamefont{J.}~\bibnamefont{Duch}} \bibnamefont{and}
  \bibinfo{author}{\bibfnamefont{A.}~\bibnamefont{Arenas}},
  \bibinfo{journal}{Physical Review E} \textbf{\bibinfo{volume}{72}},
  \bibinfo{pages}{027104} (\bibinfo{year}{2005}).

\bibitem[{\citenamefont{Pons and Latapy}(2005)}]{Pons05:CIS}
\bibinfo{author}{\bibfnamefont{P.}~\bibnamefont{Pons}} \bibnamefont{and}
  \bibinfo{author}{\bibfnamefont{M.}~\bibnamefont{Latapy}},
  \bibinfo{journal}{Computer and Information Sciences-ISCIS 2005} pp.
  \bibinfo{pages}{284--293} (\bibinfo{year}{2005}).

\bibitem[{\citenamefont{Danon et~al.}(2005)\citenamefont{Danon, Duch, Arenas,
  and D\'{\i}az-Guilera}}]{Danon:2005}
\bibinfo{author}{\bibfnamefont{L.}~\bibnamefont{Danon}},
  \bibinfo{author}{\bibfnamefont{J.}~\bibnamefont{Duch}},
  \bibinfo{author}{\bibfnamefont{A.}~\bibnamefont{Arenas}}, \bibnamefont{and}
  \bibinfo{author}{\bibfnamefont{A.}~\bibnamefont{D\'{\i}az-Guilera}},
  \bibinfo{journal}{Journal of Statistical Mechanics: Theory and Experiment} p.
  \bibinfo{pages}{P09008} (\bibinfo{year}{2005}).

\bibitem[{\citenamefont{Wolberg et~al.}(1994)\citenamefont{Wolberg, Street, and
  Mangasarian}}]{Wolberg94}
\bibinfo{author}{\bibfnamefont{W.}~\bibnamefont{Wolberg}},
  \bibinfo{author}{\bibfnamefont{W.}~\bibnamefont{Street}}, \bibnamefont{and}
  \bibinfo{author}{\bibfnamefont{O.}~\bibnamefont{Mangasarian}},
  \bibinfo{journal}{Cancer Letters} \textbf{\bibinfo{volume}{77}},
  \bibinfo{pages}{163} (\bibinfo{year}{1994}), ISSN \bibinfo{issn}{0304-3835}.

\bibitem[{\citenamefont{Witten and Frank}(2005)}]{Witten05}
\bibinfo{author}{\bibfnamefont{I.}~\bibnamefont{Witten}} \bibnamefont{and}
  \bibinfo{author}{\bibfnamefont{E.}~\bibnamefont{Frank}},
  \emph{\bibinfo{title}{{Data Mining: Practical machine learning tools and
  techniques}}} (\bibinfo{publisher}{Morgan Kaufmann Pub},
  \bibinfo{year}{2005}), ISBN \bibinfo{isbn}{0120884070}.

\end{thebibliography}

\end{document}